\documentclass[aps,pre,10pt,notitlepage,nofootinbib]{revtex4-2}

\usepackage{graphicx,amsmath,amssymb}
\usepackage{hyperref}
\hypersetup{colorlinks=true,linkcolor=blue}

\newcommand{\ells}{\xi}
\newcommand{\pd}{\partial}
\newcommand{\qhat}{\hat{q}}
\newcommand{\rhat}{\hat{r}}
\newcommand{\tenG}{\mathbb{G}}
\newcommand{\tenS}{\mathbb{S}}
\newcommand{\tensigma}{\boldsymbol{\sigma}}
\newcommand{\vecf}{{\bf f}}
\newcommand{\vecF}{{\bf F}}
\newcommand{\vecnabla}{\boldsymbol{\nabla}}
\newcommand{\vecq}{{\bf q}}
\newcommand{\vecr}{{\bf r}}
\newcommand{\vecv}{{\bf v}}

\begin{document}

\title{Model-free hydrodynamic theory of the colloidal glass transition}

\author{Haim Diamant}

\affiliation{School of Chemistry and Center for Physics and Chemistry
  of Living Systems, Tel Aviv University, Tel Aviv 6997801, Israel}

\date{November 2024}

\begin{abstract}
    We present a phenomenological, model-free theory for the
    large-distance hydrodynamic response of a viscous fluid hosting
    colloidal particles. The flow of the host fluid is affected by the
    presence of the particles, thus reflecting their liquid or solid
    state. On the liquid side of the glass transition we identify a
    dynamic length scale $\ell$ beyond which the host fluid's response
    is that of a viscous fluid with increased effective viscosity
    $\eta$. As the glass transition is approached, $\ell$ increases
    indefinitely as $\ell \sim \eta^{1/2}$. At the transition the
    large-distance response changes qualitatively, marking the host
    fluid's loss of translation invariance. On the solid side of the
    transition we identify another dynamic length, $\ell_s$, beyond
    which the host fluid responds as a viscous fluid of increased
    effective viscosity $\eta^+$, confined in a porous medium of
    effective pore size $\ell_s$. As the glass transition is
    approached from the solid side, $\ell_s$ grows indefinitely, more
    sharply than $(\eta^+)^{1/2}$. We discuss various experimental
    implications of the results and their possible relation to
    microscopic theories of the colloidal glass transition.
\end{abstract}

\maketitle

\section{Introduction}

A suspension of collidal particles in a host fluid solidifies above a
certain density or below a certain temperature. The solid may be
ordered (a colloidal crystal) or amorphous (a colloidal glass),
depending on intricate factors such as small polydispersity of
particle size or weak external stresses~\cite{Pusey1986,ChengChaikin2001}. Colloidal glasses
have been used as convenient model systems to study the elusive glass
transition~\cite{Pusey1986,WeeksScience2000,ChengChaikin2001,Conrad2006,Cipelleti2011,HunterWeeks2012,Gokhale2016,Dauchot2023}. The main advantage over ordinary glasses is that
the constituents, the colloidal particles, can be optically visualized
and tracked.

It is unclear whether the existence of a host fluid affects the
essential features of the glass transition~\cite{Conrad2006,Tateno2019,Furukawa2010}. This question is
tied to the fundamental issue of whether the mechanism underlying the
transition is structural or dynamic~\cite{BookDynamicHeterogeneities,BookStructuralGlasses}. In the present work we
turn the spotlight to the host fluid. We investigate changes in its
hydrodynamic response to a localized force, caused by the
solidification of the suspended particles.

In Sec.~\ref{sec:generic} we construct, based on symmetry arguments,
an expression for the hydrodynamic response of a generic complex
fluid. The response is captured by a set of phenomenological
coefficients, one of which serves as an order parameter for the glass
transition, i.e., it vanishes in the suspension's liquid state and is
nonzero in its solid state. We show that this critical behavior
reflects the breaking of translation invariance of the host fluid upon
the suspension's solidification. The other coefficients are found to
be related to the suspension's global viscosity and to dynamic
characteristic length scales. We then use this representation of the
host fluid's response to infer properties of the glass transition as
it is approached from the liquid side (Sec.~\ref{sec:liquid}) and the
solid side (Sec.~\ref{sec:solid}). In Sec.~\ref{sec:discussion} we
summarize the predictions of the theory, discuss their experimental
implications, and suggest possible relations between the results and
microscopic theories of the colloidal glass transition. In the
Appendix we demonstrate the emergence of the phenomenological coefficients in a simplistic microscopic model.

\section{Generic hydrodynamic response of a complex fluid}
\label{sec:generic}

We consider a suspension containing a volume fraction $\phi$ of
particles of typical size $a$ and otherwise arbitrary character, in a
fluid of viscosity $\eta_0$. In the absence of particles, the steady
flow caused by a force density $\vecf(\vecr)$ satisfies the following
equations \cite{HappelBrennerBook},
\begin{eqnarray}
    -\nabla p + \eta_0 \nabla^2\vecv + \vecf(\vecr) &=& 0, 
    \label{Stokes}\\
    \nabla\cdot\vecv &=& 0,
    \label{incompress}
\end{eqnarray}
where $\vecv(\vecr)$ is the local flow velocity and $p(\vecr)$ the
local pressure.  The Stokes equation~(\ref{Stokes}) arises from the
fluid's momentum conservation, $\nabla\cdot\tensigma+\vecf=0$, where
$\sigma_{ij}=-p\delta_{ij}+\eta_0(\pd_i v_j+\pd_j v_i)$ is the
momentum flux (stress tensor), and $\vecf$ the density of momentum
sources (force density). Accordingly, Eq.~(\ref{Stokes}) is invariant to the addition
of a uniform flow velocity, $\vecv(\vecr)\rightarrow\vecv(\vecr)+\vecv_0$,
reflecting the fluid's Galilean
invariance. Equation~(\ref{incompress}) accounts for the fluid's mass
conservation in the limit of an incompressible fluid.

In Fourier space Eqs.~(\ref{Stokes}) and (\ref{incompress}) take the form
\begin{eqnarray}
    -i\vecq p - \eta_0 q^2 \vecv + \vecf(\vecq) &=& 0, 
    \label{StokesQ}\\
    \vecq\cdot\vecv &=& 0.
    \label{incompressQ}
\end{eqnarray}
Their solution for a point force,
$\vecf(\vecr)=\vecF\delta(\vecr)\rightarrow \vecf(\vecq)=\vecF$, is
$\vecv(\vecq)=\tenG^{l,0}(\vecq)\cdot\vecF$, where
\begin{equation}
  G^{l,0}_{ij}(\vecq) = \frac{1}{\eta_0 q^2} \left( \delta_{ij} - \qhat_i
  \qhat_j \right).
\label{OseenQ}
\end{equation}
The pole $q=0$ in the prefactor reflects momentum conservation as it
arises from the conservation equation $i\vecq\cdot\tensigma+\vecf =
0$. (The pole is double due to the relation between stress and strain
rate, $\tensigma\sim i\vecq\vecv$.) The term in parentheses reflects
mass conservation as it ensures the incompressibility condition,
$\vecq\cdot\tenG = 0$. Inversion of Eq.~(\ref{OseenQ}) to real space
gives the Oseen tensor \cite{HappelBrennerBook},
\begin{equation}
    G^{l,0}_{ij}(\vecr) = \frac{1}{8\pi\eta_0 r}
    \left( \delta_{ij} + \rhat_i \rhat_j \right).
\label{Oseen}
\end{equation}

If the fluid does not conserve momentum, for example, because of
friction with immobile obstacles, Galilean invariance is broken. The
simplest way to account for it is to introduce a ``momentum-leaking" friction
term into the momentum balance equation, $\nabla\cdot\tensigma -
\Gamma\vecv+\vecf=0$, where $\Gamma$ is the leakage rate (friction coefficient). For later
convenience we rewrite $\Gamma=(\eta_0/a^2)A$ where $A$ is a
dimensionless constant. With this addition, Eq.~(\ref{Stokes}) turns
into the Brinkman equation~\cite{Brinkman1949,Durlofsky1987},
\begin{equation}
    -\nabla p + \eta_0 [\nabla^2\vecv - (A/a^2)\vecv] + \vecf(\vecr) = 0,
\label{Brinkman}
\end{equation}
which is no longer invariant under $\vecv(\vecr)\rightarrow\vecv(\vecr)+\vecv_0$.
This equation has been widely used to describe viscous flow
through porous media. The characteristic length,
\begin{equation}
    \ells = a A^{-1/2},
\end{equation}
is associated with the typical pore size. 

In Fourier space Eq.~(\ref{Brinkman}) turns into
\begin{equation}
    -i\vecq p - \eta_0 (\ells^{-2} + q^2)\vecv + \vecf(\vecq) = 0.
\label{BrinkmanQ}
\end{equation}
For $\vecf(\vecr)=\vecF\delta(\vecr)$, the solution of this equation
together with Eq.~(\ref{incompressQ}) is
$\vecv(\vecq)=\tenG^{s,0}(\vecq)\cdot\vecF$, with
\begin{equation}
    G^{s,0}_{ij}(\vecq) = \frac{1}{\eta_0(\ells^{-2} + q^2)}
    \left( \delta_{ij} - \qhat_i \qhat_j \right).
\label{OseenQS}
\end{equation}
The $q=0$ pole in the prefactor has disappeared, indicating the
breaking of momentum conservation (i.e., translation
invariance). Looking at the first factor only, one expects
$\tenG^s(\vecr)$ in real space to decay exponentially with
$r/\ells$. However, the mass-conservation term still contains a pole
in the form $-\qhat_i\qhat_j=-q_iq_j/q^2$, which transforms to
\begin{equation}
  -\frac{q_iq_j}{q^2} \rightarrow \pd_i\pd_j \frac{1}{4\pi r} =
  -\frac{1}{4\pi r^3} (\delta_{ij} - 3\rhat_i\rhat_j).
\end{equation}
Thus, this term gives rise to a long-range $1/r^3$ decay [albeit
shorter than the $1/r$ decay of $\tenG^l(\vecr)$]. It has the angular
shape of a dipolar flow as it originates from the effective mass dipole
created by the point force over the ``pore'' size $\ells$
\cite{DiamantIJC2007,DiamantJPSJ2009}. Full inversion of Eq.~(\ref{OseenQS}) gives
\begin{eqnarray}
    G^{s,0}_{ij}(\vecr) = -\frac{1}{4\pi\eta_0\ells} && \left\{
    \frac{1}{r_s^3} \left( \delta_{ij} - 3\rhat_{s,i}\rhat_{s,j} \right) \right. 
    \label{OseenS} \\
    &&- \left. \frac{e^{-r_s}}{r_s^3} \left[ (r_s^2 + r_s + 1) \delta_{ij} - (r_s^2 + 3r_s + 3) \rhat_{s,i}\rhat_{s,j} \right] \right\},\ \ \ \ \vecr_s = \vecr/\ells. \nonumber
\end{eqnarray}
Over scales larger than the pore size, $r\gg\ells$, the asymptotic response has the expected $1/r^3$
decay, with only exponentially small corrections. For $r\ll\ells$
Eq.~(\ref{OseenS}) reduces to the fluid response, Eq.~(\ref{Oseen}).

\begin{figure}
    \centering
    \includegraphics[width=0.7\textwidth]{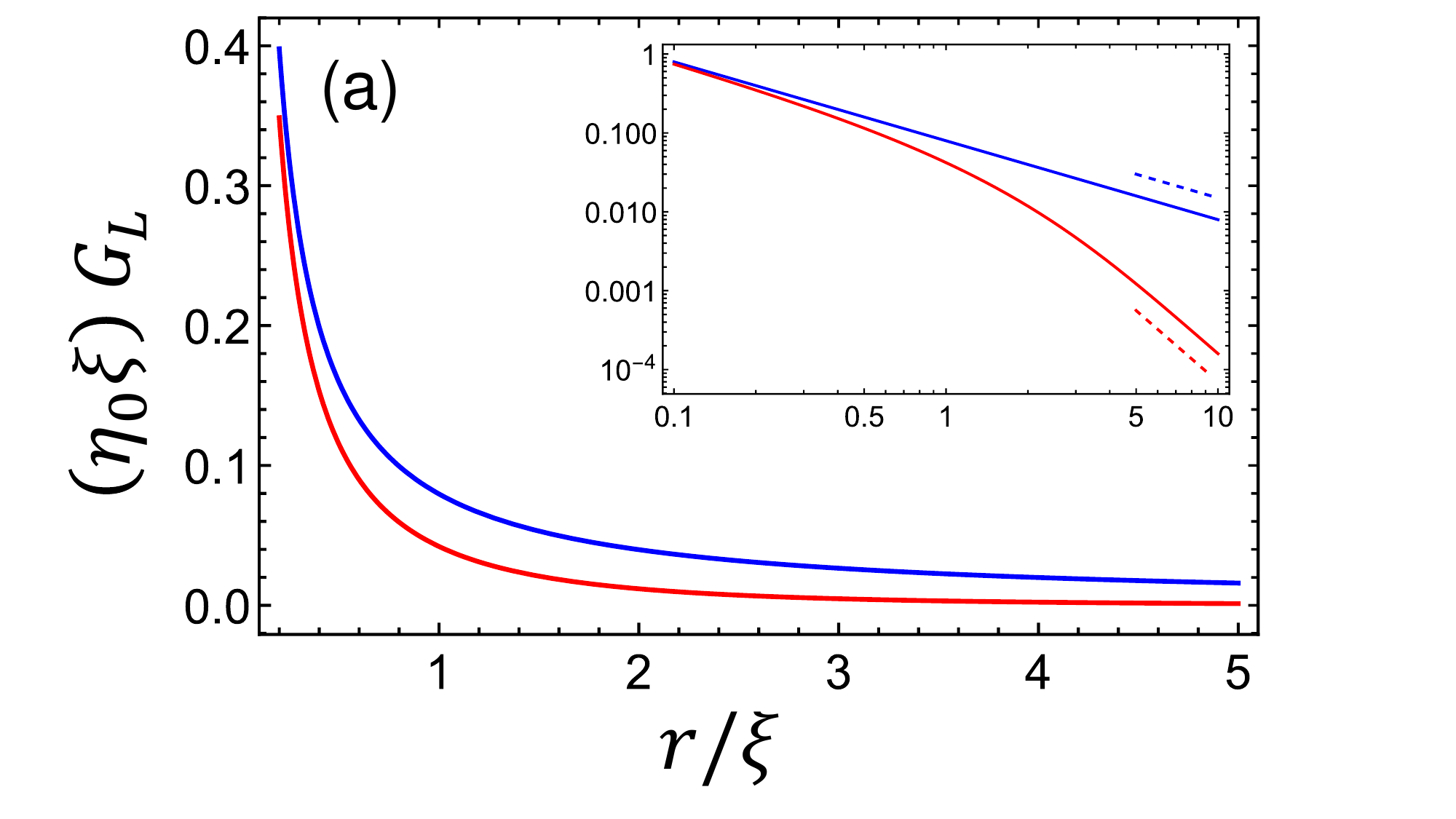}
    \includegraphics[width=0.7\textwidth]{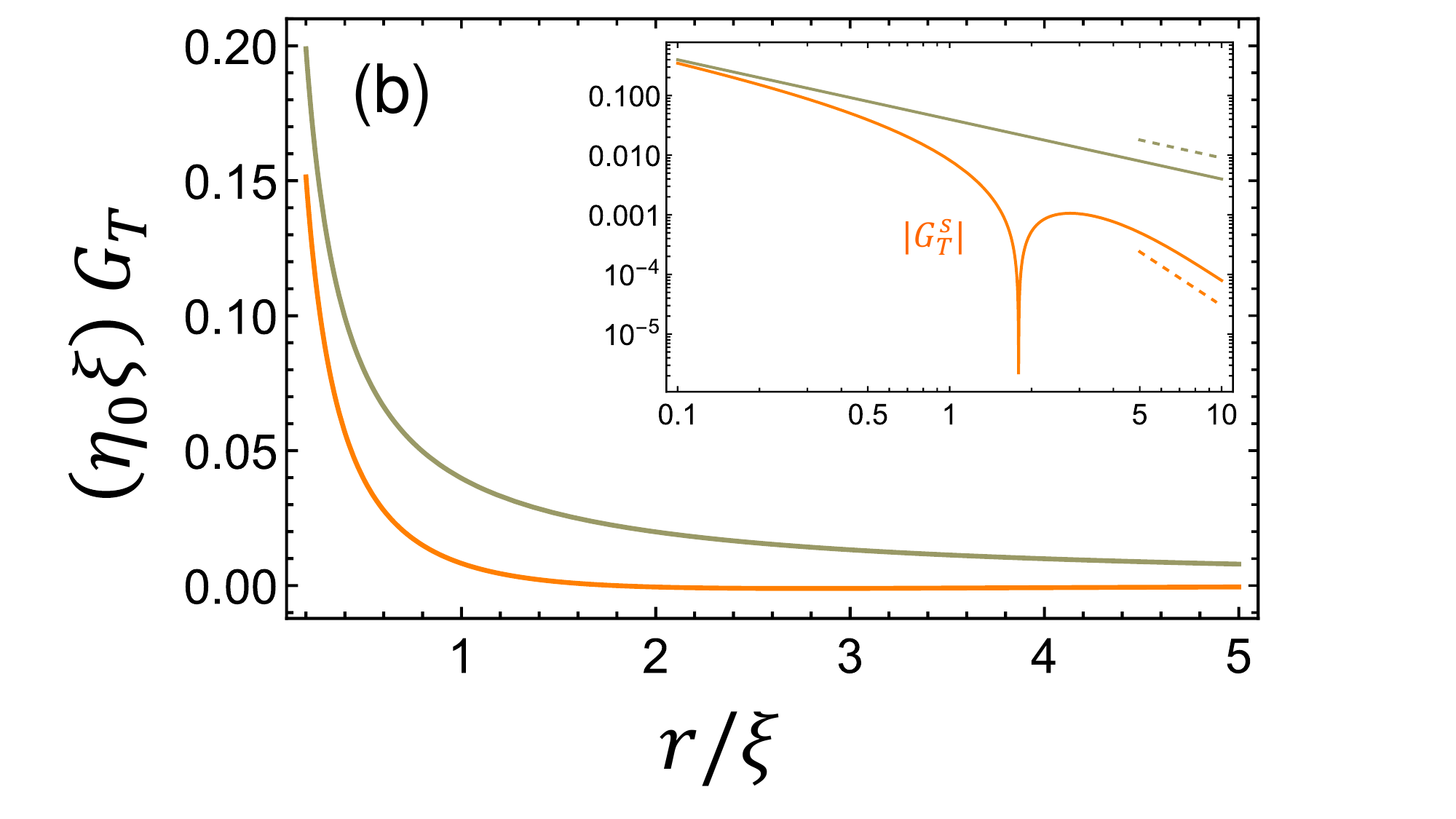}
    \caption{
      The longitudinal (a) and transverse (b) hydrodynamic
  responses as a function of distance from an applied point force, for
  a momentum-conserving fluid ($G^{l,0}$, upper curves) and a
  momentum-nonconserving fluid ($G^{s,0}$, lower curves). The insets
  show the responses on a logarithmic scale, with the dashed lines
  indicating $1/r$ and $1/r^3$ decays. The responses are
  scaled by $(\eta_0\ells)^{-1}$ and the distance by $\ells$. For
  $r\ll\ells$, $G^{l,0}_{L.T}$ and $G^{s,0}_{L,T}$ coincide. The transverse response
  $G^{s,0}_T$ becomes negative for $r/\ells\gtrsim 1$ [panel (b) inset].}
\label{fig:GLT}
\end{figure}

The response tensor $\tenG$ of an isotropic system can conveniently be represented by two
scalars, the longitudinal and
transverse responses,
\begin{equation}
  G_L(r)=G_{xx}(r{\bf\hat{x}}),\ \ \ \ 
  G_T(r)=G_{xx}(r{\bf\hat{y}}).
\end{equation}
Figure~\ref{fig:GLT} shows these response modes for the
momentum-conserving fluid [Stokes flow, Eq.~(\ref{Oseen})] and the
momentum-nonconserving one [Brinkman flow, Eq.~(\ref{OseenS})]. In
addition to the different asymptotic decay, $1/r$ and $1/r^3$,
respectively, the momentum-nonconserving fluid has another
distinctive feature. Its asymptotic transverse response is
negative. This follows from the dipolar shape of the first term in
Eq.~(\ref{OseenS}), as mentioned above.

As concerns complex fluids, such as colloidal suspensions and polymer
networks, the two hydrodynamic responses discussed above correspond to
two limits where the flow is of the host fluid alone. Equations
(\ref{OseenQ}) and (\ref{Oseen}) give the response of the
particle-free fluid, i.e., the limit of vanishing volume fraction of
particles. Equations (\ref{OseenQS}) and (\ref{OseenS}) give the
response in the case where the host fluid flows past an immobile
matrix. As these two extremes are separated by a symmetry breaking,
their asymptotic spatial decays are distinct. The translational
symmetry breaking is captured by the coefficient $A$, which serves as
an order parameter, becoming nonzero when the symmetry is
broken. Moreover, the asymptotic $1/r$ and $1/r^3$ power laws arise
from conservation laws of momentum and mass, respectively. Hence, they must
hold for the host fluid's response in any liquid and solid suspension,
respectively~\cite{DiamantIJC2007,DiamantJPSJ2009}.

In the present work we would like to treat suspensions away from these
two limits, in particular, in the vicinity of the symmetry breaking. As the
volume fraction $\phi$ of a suspension is increased from zero, length
scales emerge which characterize the structure of the suspended
particles. A natural way to introduce these length scales into the
coarse-grained hydrodynamic response by the following extension of Eqs.~(\ref{StokesQ}) and (\ref{BrinkmanQ}),
\begin{equation}
    -i\vecq p + \eta_0 [-A/a^2 + B(iq)^2 + Ca^2 (iq)^4 + Da^4(iq)^6 +
      \ldots ] \vecv + \vecf(\vecq) = 0.
\label{Stokesgeneric}
\end{equation}
where $A(\phi), B(\phi), C(\phi), D(\phi),\ldots$ are dimensionless
coefficients which depend on the volume
fraction (and in general also the temperature).\footnote{\label{ft:visco} 
In
this work we address the suspension's purely viscous response, i.e.,
the limit of vanishing frequency. More generally, for a viscoelastic
suspension or gel, these coefficients will be frequency-dependent.}
Each additional term accounts for the effect of a higher-order
velocity gradient, i.e., a smaller (yet still large compared to $a$) length
scale.\footnote{
An equivalent way to
view Eq.~(\ref{Stokesgeneric}) is to consider the entire factor
multiplying $\vecv$ as a $q$-dependent (and possibly also
frequency-dependent) viscosity~\cite{Grosberg2016}.} The solution of
Eq.~(\ref{Stokesgeneric}) for a point force
$\vecf(\vecr)=\vecF\delta(\vecr)$, together with the
incompressibility condition (\ref{incompressQ}), is
$\vecv=\tenG\cdot\vecF$ with
\begin{equation}
    G_{ij}(\vecq) = \frac{1}{\eta_0( A/a^2 + Bq^2 - Ca^2q^4 + Da^4q^6
      - \cdots)} \left( \delta_{ij} - \qhat_i\qhat_j \right).
\label{Ggeneric}
\end{equation}

The Ansatz in Eq.~(\ref{Ggeneric}) concerning the hydrodynamic
response of the host fluid is our starting point, from which the
rest of the results inevitably follow. The generic expansion in
Eqs.~(\ref{Stokesgeneric}) and (\ref{Ggeneric}) assumes only that the
host fluid remains rotation- and inversion-symmetric, and
incompressible, thus making the theory model-free. The details of a
specific system are absorbed in the unspecified dependencies of the
phenomenological coefficients $A,B,C,D,\ldots$ on
$\phi$ (and temperature).$^{\ref{ft:visco}}$ In the Appendix we obtain the leading
coefficients explicitly for a simplistic example.


\section{Liquid side of the transition}
\label{sec:liquid}

In a liquid suspension, where all constituents are mobile and momentum
is conserved, $A=0$ exactly. We rewrite Eq.~(\ref{Ggeneric}) as
\begin{equation}
    G_{ij}(\vecq) = \frac{1}{\eta_0 B [q^2 - (C/B)a^2q^4 + (D/B)a^4q^6
        - \ldots]} \left( \delta_{ij} - \qhat_i\qhat_j \right),
\label{Gliquid}
\end{equation}
which can be decomposed as
\begin{eqnarray}
    G^l_{ij}(\vecq) &=& \left[ \frac{1}{\eta q^2} +
      \frac{1-(D/C)a^2 q^2 + \ldots}{\eta( \ell^{-2} - q^2 + (D/C)a^2q^4 - \ldots )}
      \right] \left( \delta_{ij} - \qhat_i\qhat_j \right),
    \label{OseenQeff}\\
    \eta(\phi) &=& \eta_0 B(\phi),\\
    \ell(\phi) &=& a [C(\phi)/B(\phi)]^{1/2}.
\end{eqnarray}
The first term in Eq.~(\ref{OseenQeff}) coincides with the response of
a momentum-conserving fluid, Eq.~(\ref{OseenQ}), except that the
particle-free viscosity $\eta_0$ has been replaced by
$\eta(\phi)=B(\phi)\eta_0$. Thus the coefficient $B(\phi)$ relates to
the effective viscosity of the suspension. The second term in
Eq.~(\ref{OseenQeff}) contains a pole structure that depends on the
roots of the polynomial $(\ell^{-2}-q^2 +(D/C)a^2q^4 - \ldots)$. Each
of these roots will give rise in real space to an exponential term
with a certain decay length (and possibly also oscillation wavelength
if the root is complex), reflecting the structure of the
suspension. However, for any finite $\ell$, the polynomial does not
have a root equal to zero. The only $q=0$ root comes, as in the response of
a momentum-nonconserving fluid [Eq.~(\ref{OseenQS})], from the dipolar
mass term, $-(\ell^2/\eta)q_iq_j/q^2$. Therefore, we can directly read
from Eqs.~(\ref{Oseen}) and (\ref{OseenS}) the large-distance response
of the liquid suspension as
\begin{equation}
  G^l_{ij}(\vecr) = \frac{1}{8\pi\eta r} \left(\delta_{ij} +
  \rhat_i\rhat_j \right) - \frac{\ell^2}{4\pi\eta r^3} \left(\delta_{ij}
  - 3\rhat_i\rhat_j \right),
  \label{Oseeneff}
\end{equation}
with exponentially small corrections.

\begin{figure}
    \centering
    \includegraphics[width=0.7\textwidth]{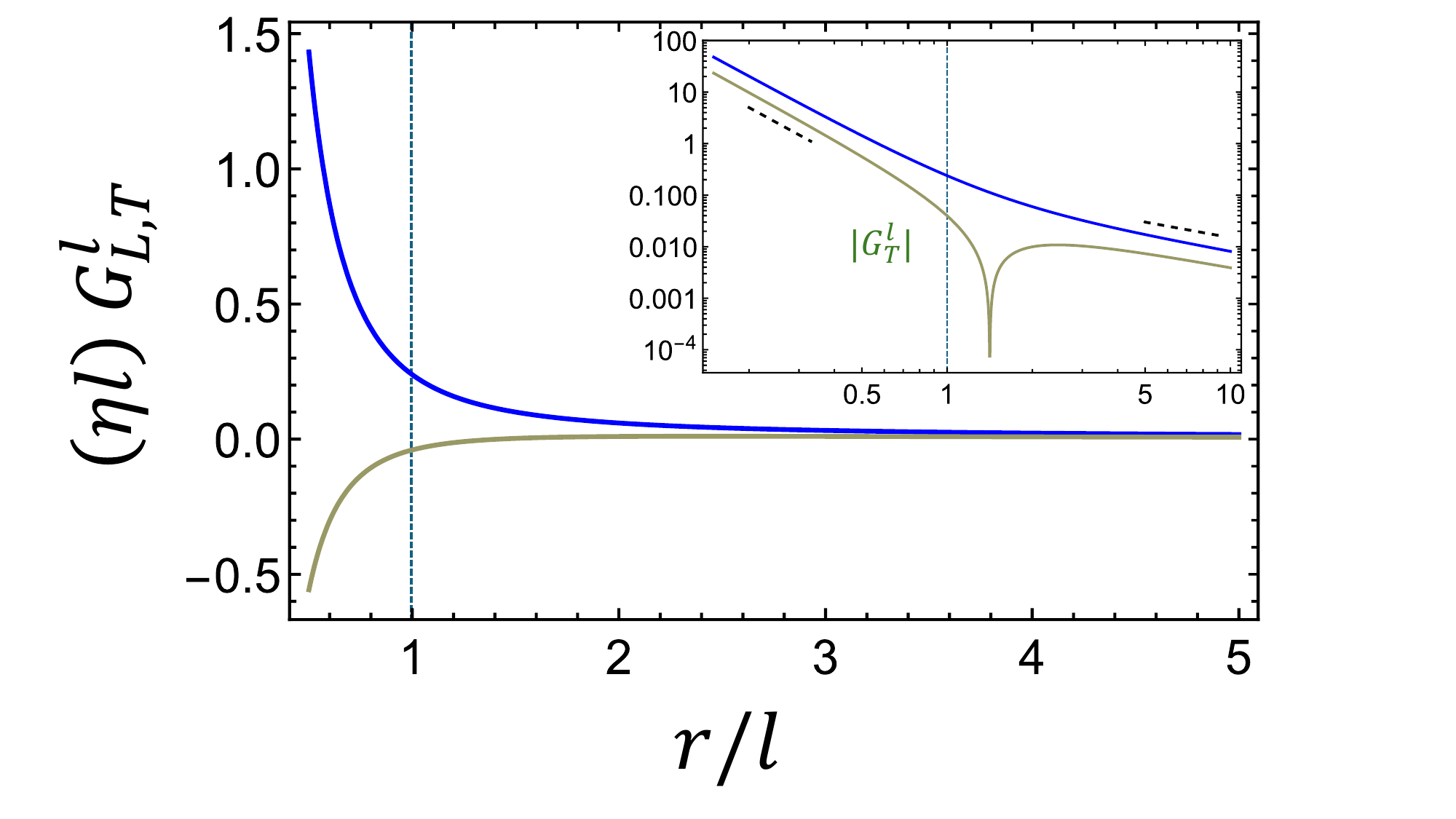}
    \caption{
      The longitudinal (upper curve) and transverse (lower curve) hydrodynamic responses as a function of distance from an
  applied point force, for a liquid suspension. The inset
  shows the responses on a logarithmic scale, with the dashed lines
  indicating $1/r$ and $1/r^3$ decays. The responses are scaled by
  $(\eta\ell)^{-1}$ and the distance by $\ell$. Below the distance
  $r\sim\ell$ (dotted vertical lines) the response crosses over from a
  $1/r$ decay to a $1/r^3$ decay, and the transverse response becomes
  negative.}
\label{fig:GLTliq}
\end{figure}

Thus the host fluid's response contains a solid-like component as
a subdominant term at large distances. It arises from the effective
mass dipole created by the point force over the intrinsic length scale
$\ell(\phi)$, similar to the one formed in a momentum-nonconserving
fluid over the length scale $\ells$ (where this term is
dominant). The distance $r\sim\ell$ marks a crossover from the
liquid-like response, $\sim 1/(\eta r)$, for $r\gtrsim\ell$, to the
solid-like one, $\sim \ell^2/(\eta r^3)$, for $r\lesssim\ell$. The
longitudinal and transverse responses arising from
Eq.~(\ref{Oseeneff}) are plotted in Fig.~\ref{fig:GLTliq}.

As the glass transition is approached, $\eta(\phi)$ grows
indefinitely, making the dominant response, $\sim 1/(\eta r)$
increasingly weak. As we have seen in Sec.~\ref{sec:generic}, however,
the momentum-nonconserving response of the host fluid
is finite in a solid. Hence, the subdominant term, $\sim \ell^2/\eta$, should
remain finite. This implies that the crossover length $\ell(\phi)$
must grow indefinitely as well,
as\footnote{
If the suspension has no
other intrinsic structural length than $a$, and no other intrinsic
time than $a^2/(\eta_0/\rho_0)$ ($\rho_0$ being the mass density of
the host fluid), then, on dimensional grounds, $\ell(\phi) \sim
(a/\eta_0)\eta(\phi)$. Equivalently, $\ell^2/\eta \sim a^2/\eta_0$.
The physical interpretation is that the strength of the mass dipole,
underlying the $1/r^3$ response, depends on the local viscosity rather 
than the suspension's large-scale viscosity
$\eta(\phi)$.}
\begin{equation}
  \ell(\phi) \sim \eta(\phi)^{1/2} \xrightarrow{\eta\rightarrow\infty} \infty.
\end{equation}

\section{Solid side of the transition}
\label{sec:solid}

\begin{figure}
    \centering
    \includegraphics[width=0.7\textwidth]{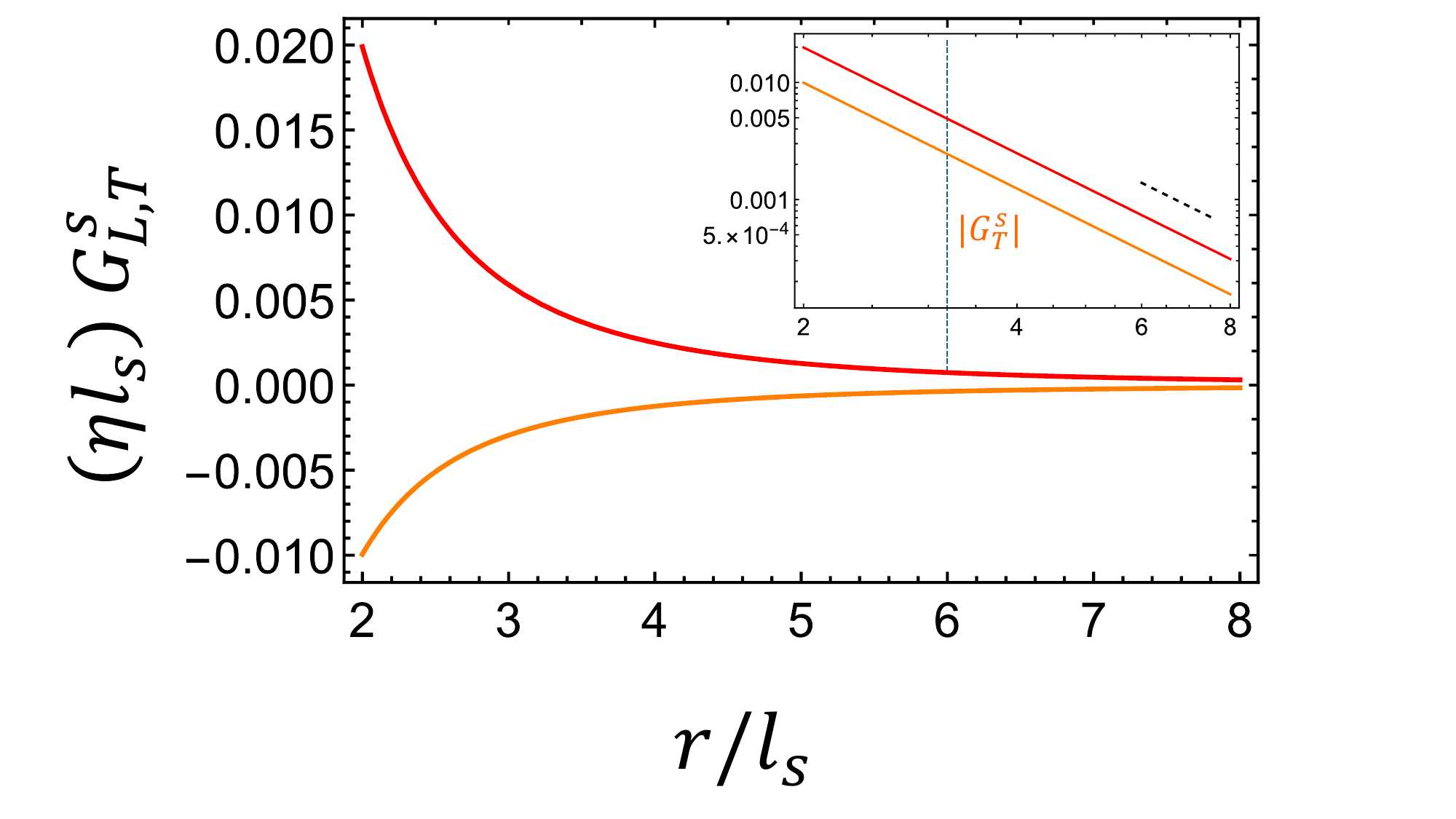}
    \caption{
      The longitudinal (upper curve) and transverse (lower
  curve) hydrodynamic responses as a function of distance from
  an applied point force, at large distances, for a solid suspension. The inset shows the responses on a logarithmic scale,
  with the dashed line indicating a $1/r^3$ decay. The responses are
  scaled by $(\eta^+\ell_s)^{-1}$ and the distance by $\ell_s$. For
  $r\gtrsim\ell_s$ the transverse response is negative.}
\label{fig:GLTsol}
\end{figure}

For a solid we have $A>0$, leading to a qualitatively
different response at large distances. We return to
Eq.~(\ref{Ggeneric}) and rewrite it now as
\begin{eqnarray}
  G^s_{ij}(\vecq) &=& \frac{1}{\eta^+ [\ell_s^{-2} + q^2 - (C^+/B^+)a^2q^4
      + (D^+/B^+)a^4q^6 - \ldots]} \left( \delta_{ij} - \qhat_i\qhat_j \right).
  \label{OseeneffQS}\\
  \eta^+(\phi) &=& \eta_0 B^+(\phi), 
  \label{etaplus}\\
  \ell_s(\phi) &=& a [B^+(\phi)/A(\phi)]^{1/2} = \ells\,
      [\eta^+(\phi)/\eta_0]^{1/2}.
  \label{ells}
\end{eqnarray}
We have added the superscript `$+$' to indicate that the coefficients
$B,C,D,\ldots$ are not supposed to have the same $\phi$-dependence
below and above the transition. As before, the detailed response
depends on the pole structure of Eq.~(\ref{OseeneffQS}), but, for any
finite $\ell_s$, the only $q=0$ pole comes again from the mass-dipole
term, $-(\ell_s^2/\eta^+)q_iq_j/q^2$. Hence, replacing $\eta_0$ by
$\eta^+$ and $\xi$ by $\ell_s$, we can directly read from
Eq.~(\ref{OseenS}) the large-distance response of the host fluid in
the solid suspension,
\begin{equation}
  G^s_{ij}(\vecr) = - \frac{\ell_s^2}{4\pi\eta^+ r^3} \left(\delta_{ij}
  - 3\rhat_i\rhat_j \right),\ \ \ \ r > \ell_s,
  \label{OseeneffS}
\end{equation}
with corrections that are exponentially small in $r/\ell_s$. The
longitudinal and transverse components of this response are shown in
Fig.~\ref{fig:GLTsol}. For $r\gtrsim\ell_s$ [the domain where
Eq.~(\ref{OseeneffS}) is valid], the transverse correlation is
negative.

Thus the host fluid in a solid suspension behaves over large scales
like a Brinkman fluid with effective viscosity $\eta^+(\phi)$ and
effective pore size $\ell_s(\phi)$. Away from the glass transition the
solid matrix should become rigid, and we expect $B^+$ to tend to $1$,
making $\eta^+$ tend to $\eta_0$ and $\ell_s$ tend to $\xi$, thus
coinciding with the Brinkman parameters for a rigid porous medium
[Eqs.~(\ref{OseenQS}) and (\ref{OseenS})]. As the glass transition is
approached from the solid side, assuming that amorphous solidification
is a continuous transition, $A(\phi)$ should continuously decrease to
zero. This is accompanied by changes of $\eta^+$ and
$\ell_s$. According to Eqs.~(\ref{etaplus}) and (\ref{ells}), $\ell_s$
must diverge at least as sharply as $A^{-1/2}$. In addition, just
above the transition, there is a single effective pore of size
$\ell_s\rightarrow\infty$, whose viscosity $\eta^+$ must diverge as
well to describe a solid. Hence, $\ell_s$ increases close to the
transition more sharply than $A^{-1/2}$,
\begin{equation}
  \ell_s(\phi) \sim A(\phi)^{-1/2} \eta^+(\phi)^{1/2}
  \xrightarrow{A\rightarrow 0,\eta^+\rightarrow\infty} \infty.
\end{equation}

\section{Discussion}
\label{sec:discussion}

We have introduced a set of phenomenological coefficients,
characterizing the hydrodynamic response of a suspension's host fluid over
decreasing (yet large) distances, and demanded that the fluid's
conservation laws be satisfied or broken. This has led to a set of
nontrivial conclusions. We begin by summarizing the predictions of the
theory.\footnote{
We refer here to the transition point in terms of a critical volume fraction $\phi_c$ for a fixed temperature. We may just as well consider a critical temperature at a fixed volume fraction (for thermal systems, i.e., not merely hard spheres).}

The solidification of the suspended particles causes a critical
behavior of their host fluid. The critical behavior is captured by
three phenomenological coefficients, $A(\phi)$, $B(\phi)$, and
$C(\phi)$, which should depend also on temperature and
be different on the two sides of the transition. The coefficient $A$
serves as an order parameter, being strictly zero in a liquid
suspension, where the host fluid is translation-invariant, and
nonzero in a solid suspension, where the translation invariance of the
host fluid is broken. The coefficient $B$ is the ratio between the
effective viscosity $\eta$ and the viscosity $\eta_0$ of the pristine
host fluid.

Below the glass transition ($A=0$) $B$ and $C$ give rise to a dynamic
length $\ell\sim(C/B)^{1/2}$. It marks the crossover of the host
fluid's response from a $1/(\eta r)$ decay at asymptotically large
distances to a ``solid-like'' $\ell^2/(\eta r^3)$ decay at smaller
distances. The latter's transverse component is negative. The dynamic
length $\ell$ increases with effective viscosity as
$\ell\sim\eta^{1/2}$ \cite{HuPRE2022}. Thus, as $\eta$ increases, the crossover occurs
at an increasingly large distance, until, at the transition, the
solid-like $1/r^3$ response takes over the whole system. If the
effective viscosity (or the relaxation time) diverges as a power law \cite{Cipelleti2011},
\begin{equation}
  \eta(\phi) \sim (\phi_c-\phi)^{-\zeta}\ \ \ \ \phi\nearrow\phi_c,
\end{equation}
with some exponent $\zeta$,
then
\begin{equation}
  \ell(\phi) \sim
  (\phi_c-\phi)^{-\zeta/2},\ \ \ \ \phi\nearrow\phi_c^-.
\end{equation}
We would like to underline the significance and usefulness of the
transverse response's different sig for $r\lesssim\ell$ and
$r\gtrsim\ell$. As long as the transverse response approaches zero
from above at very large distances, the suspension is overall liquid.
This is a clearcut experimental criterion which does not require fitting. Unlike the common tests of glassiness, such as the confined
dynamics of single particles or a non-decaying dynamic scattering
function at $q\sim 1/a$, this test offers a probe of the
suspension's {\it large-scale} fluidity or solidity.

Above the transition ($A>0$) the coefficients $A$ and $B$ give rise to
a dynamic length $\ell_s\sim(B/A)^{1/2}$. It acts as a dynamic
``pore'' size beyond which the solid matrix affects the host fluid as
a stationary porous medium. Assuming that the transition is
continuous, one expects the symmetry breaking to affect the order
parameter as
\begin{equation}
  A(\phi) \sim (\phi-\phi_c)^\beta,\ \ \ \ \phi\searrow\phi_c,
\end{equation}
with some exponent $\beta$. If, in addition, the effective pore
viscosity increases as
\begin{equation}
  \eta^+(\phi) \sim (\phi-\phi_c)^{-\zeta^+},\ \ \ \ \phi\searrow\phi_c,
\end{equation}
then the dynamic length on the solid size diverges as
\begin{equation}
  \ell_s(\phi) \sim (\phi-\phi_c)^{-(\beta + \zeta^+)/2},\ \ \ \
  \phi\searrow\phi_c.
\end{equation}

Most of the predictions concerning the approach to the glass
transition from the liquid side have been verified in recent
experiments by Egelhaaf and coworkers~\cite{Laermann2025}: (a) the asymptotic
$1/(\eta r)$ response with $\eta(\phi)$ as the suspension's effective
viscosity; (b) the crossover to the $\ell^2/(\eta r^3)$ response at
smaller distances, with a negative transverse component; (c) the
sharply increasing length $\ell(\phi)$, extracted from the crossover,
as the transition is approached. The experiments used confocal
microscopy to track the correlated Brownian motions of pairs of tracer
particles in a suspension of much larger particles. The measured
displacement correlations of the tracers are proportional to the
responses calculated here owing to the fluctuation-dissipation
theorem~\cite{CrockerPRL2000}. The experiments were not able to check the predictions
concerning the solid suspension. For all the studied volume fractions the suspension was found to be a liquid \cite{BrambillaPRL2009} over large distances, using the transverse-response criterion
mentioned above.

The results of the present work have been obtained from the
phenomenological Ansatz of Eq.~(\ref{Ggeneric}) and do not depend on a
specific microscopic (particle-based) model. This has pros and
cons. One advantage is that the predictions should apply to any
complex fluid containing a molecular host fluid. Indeed, the
crossover from the asymptotic $1/r$ response to the $1/r^3$ one, with
the associated crossover length $\ell$, were observed not only in colloidal
suspensions \cite{Laermann2025} but also in entangled networks of semiflexible
polymers~\cite{Sonn-SegevPRL2014,Sonn-SegevSM2014}. Thus our findings should be relevant for transitions in other complex fluids such as fiber networks \cite{LicupPNAS2015} and the cytoplasm~\cite{Zhou2009}. Another advantage is that the results follow from a
minimum set of assumptions and can be used to test the validity of
more detailed theories.

This work's obvious shortcoming is the lack of information concerning the particle-level mechanisms behind the transition. Microscopic models (e.g., for hard sphere) should be able to provide explicit expressions for the
coefficients $A,B,C,D,\ldots$. Nevertheless, the phenomenological description of the
transition obtained here suggests a relation to observations concerning dynamic
heterogeneities~\cite{BookDynamicHeterogeneities}. On the liquid side, the solid-resembling
dynamic response for $r\lesssim\ell$ may be associated with
clusters \cite{KroyPRL2004,Conrad2006} of typical size $\ell$ in an overall liquid suspension. On the
solid side, the growing effective ``pores'' of typical size $\ell_s$
may be associated with liquid ``soft spots'' in an overall
solid suspension~\cite{AharonovEPL2007}. In addition, the order parameter $A(\phi)$ might be
interpreted in terms of the fraction of immobile particles (see the Appendix).

In conclusion, the phenomenological theory developed above reveals symmetry-related constraints imposed on the host fluid below and and above the colloidal glass transition. It highlights the transition's critical effect on the
dynamics of the host fluid and what can be learned from this effect on the
transition. The theory thus sets colloidal glasses apart from
molecular glasses. The predictions concerning the solid side of the
transition are yet to be checked experimentally. The correspondence between the hydrodynamic theory and 
particle-based theories calls for further investigation.

\acknowledgments 
The theory was developed to accompany experiments by
Prof.\ Dr.\ Stefan U.\ Egelhaaf and his group~\cite{Laermann2025}. This article is
dedicated to the dear memory of Stefan. I am grateful to Patrick
Laermann, Manuel Escobedo-S\'anchez, Yael Roichman, and Ivo Buttinoni, who
participated in that joint project. I thank Paddy Royall for
illuminating talks. The research was supported by the German-Israeli Foundation (GIF Grant no.\ I-1345-303.10-2016) and the Israel Science Foundation (ISF Grant no.\ 1611/24).

\appendix*

\section{Particle-based model}

This Appendix demonstrates the emergence of the phenomenological
coefficients introduced in Sec.~\ref{sec:generic} in a specific,
simplistic example. The model consists of hard spheres of radius $a$
and number density $c$ suspended in a viscous host fluid of viscosity
$\eta_0$. A fraction $\alpha$ of the particles are restricted in their
motion to the vicinity of a certain position, whereas the rest, a
fraction of $1-\alpha$, are free to move. As regards the hydrodynamics
of the host fluid, the former particles can resist a force whereas the latter particles
are force- and torque-free. For simplicity we assume that the
restricted particles are torque-free. The extension to particles whose
rotation is restricted as well is straightforward.
Our aim is to calculate the large-distance hydrodynamic response of
the host fluid to leading order in $c$. Practically, it is inconsistent to
assume that particles are spontaneously restricted (caged) in a dilute
suspension. This, and the neglect of positional correlations, limits the model to demonstrative purposes. Still, one can come up with physical scenarios where
the situation described above is achievable, e.g., by applying optical
traps to immobilize a fraction of particles in a dilute suspension.

We consider a point force, $\vecF\delta(\vecr)$, applied to the host
fluid at the origin. To zeroth order in $c$, the force creates the
flow velocity [see Eq.~(\ref{Oseen})]
\begin{equation}
  \vecv^0(\vecr) = \tenG^0(\vecr)\cdot\vecF,\ \ \ \
  G^0_{ij}(\vecr) = \frac{1}{8\pi\eta_0 r}
  \left( \delta_{ij} + \rhat_i\rhat_j \right).
\label{v0}
\end{equation}
Next we consider a particle located at $\vecr'$ and experiencing the
flow (\ref{v0}). A restricted particle responds by
exerting a force given by Fax\'en's first law~\cite{HappelBrennerBook},
\begin{equation}
  \vecF^1(\vecr') = -6\pi\eta_0 a \left[ 1 + (a^2/6) \nabla^2 \right]
  \vecv^0(\vecr') =
  -6\pi\eta_0 a \left[ 1 + (a^2/6) \nabla^2 \right]
  \tenG^0(\vecr')\cdot\vecF,
\label{F1}
\end{equation}
and a symmetric force dipole (stresslet) given by Fax\'en's third
law~\cite{HappelBrennerBook},
\begin{eqnarray}
  \tenS^1(\vecr') &=& \frac{10\pi}{3} \eta_0 a^3 \left[ 1 + (a^2/10)
    \nabla^2 \right] \left[ \vecnabla \vecv^0(\vecr') +
    (\vecnabla\vecv^0(\vecr'))^{\rm T} \right] \nonumber\\
  &=& \frac{10\pi}{3} \eta_0 a^3 \left[ 1 + (a^2/10)
    \nabla^2 \right] \left[ \vecnabla \tenG^0(\vecr')\cdot\vecF +
    (\vecnabla\tenG^0(\vecr'))^{\rm T}\cdot\vecF \right],
\label{S1}
\end{eqnarray}
assuming that it is torque-free.  A force- and torque-free particle
responds by $\tenS^1$ alone. For a rigid sphere there are no higher
force moments~\cite{HappelBrennerBook}. The force introduced by a restricted particle
at $\vecr'$ changes the flow at $\vecr$ by
\begin{equation}
  \vecv^1_F(\vecr,\vecr') = \tenG^0(\vecr-\vecr') \cdot \vecF^1(\vecr').
\label{v1F}
\end{equation}
The force dipole introduced by either type of particle at $\vecr'$ changes the flow at $\vecr$ by
\begin{equation}
  \vecv^1_S(\vecr,\vecr') = \vecnabla\tenG^0(\vecr-\vecr')
  \cdot \tenS^1(\vecr').
\label{v1S}
\end{equation}

We now add the contributions to the flow at $\vecr$ from all the
particles. To linear order in $c$ the probability densities of finding
a restricted or free particle at a certain position are uniform and
equal to $\alpha c$ and $(1-\alpha)c$, respectively. Thus the average
flow velocity at $\vecr$ is given by
\begin{equation}
  \langle \vecv(\vecr) \rangle = \vecv^0(\vecr) + c \int d\vecr' \left[
    \alpha \vecv^1_F(\vecr,\vecr') + \vecv^1_S(\vecr,\vecr') \right].
\label{vav1}
\end{equation}
Transforming Eqs.~(\ref{F1})--(\ref{vav1}) to Fourier space, with
\begin{equation}
  G^0_{ij}(\vecq) = \frac{1}{\eta_0 q^2} \left( \delta_{ij} - \qhat_i\qhat_j
  \right),
\end{equation}
and making use of the convolution theorem, we find
\begin{eqnarray}
  \langle \vecv(\vecq) \rangle &=& \tenG(\vecq) \cdot \vecF \nonumber\\
  \tenG(\vecq) &=& \tenG^0(\vecq) -
  \frac{\pi a c}{q^2}
  \left[ 6\alpha + (10/3-\alpha) a^2q^2 - (1/3) a^4q^4 \right] \tenG^0(\vecq).
\end{eqnarray}
This can be rearranged, to leading order in $c$, as
\begin{equation}
  G_{ij}(\vecq) = \frac{1}{\eta_0}\, \frac{1}{(9/2)\alpha\phi/a^2 +
    [1 + (5/2)\phi (1 - 3\alpha/10)] q^2 - (\phi a^2/4) q^4} \left( 
    \delta_{ij} - \qhat_i\qhat_j \right).
\end{equation}
where we have replaced $\phi=(4\pi a^3/3)c$. Comparing with
Eq.~(\ref{Ggeneric}) we read
\begin{equation}
  A = \frac{9}{2} \alpha\phi,\ \ \ \ 
  B = 1 + \frac{5}{2} \phi \left( 1 - \frac{3}{10}\alpha \right),\ \ \ \
  C = \frac{1}{4}\phi.
\end{equation}

Thus the order parameter $A$ is proportional to the fraction
$\alpha$ of restricted particles, i.e., as expected, it vanishes for a
liquid suspension where all particles are free. The effective
viscosity $\eta=\eta_0 B$ is appropriately positive. For
$\alpha=0$ it coincides with the classical result~\cite{EinsteinBook},
$\eta=\eta_0[1+(5/2)\phi]$. For $\alpha>0$ it decreases with $\alpha$, which may seem non-intuitive but is the appropriate trend.
As discussed in Sec.~\ref{sec:solid}, when
we get further away from the transition on the solid side, the
effective pore size $\ell_s \sim (B/A)^{1/2}$ decreases toward $\xi$
and its effective viscosity decreases toward $\eta_0$. Because of the dependence on $\alpha(\phi)$, $B$ has a different $\phi$-dependence on the two sides of the transition.

\end{document}